\documentstyle[twoside,fleqn,espcrc2]{article}
\input epsf
\newcommand{\bc}{\begin{center}}
\newcommand{\ec}{\end{center}}
\newcommand{\beqn}{\begin{equation}}
\newcommand{\eeqn}{\end{equation}}
\newcommand{\barr}{\begin{eqnarray}}
\newcommand{\earr}{\end{eqnarray}}
\def\del{\partial}

\def\eg {{\it e.g}. }
\def\ie {{\it i.e}. }

\def\tr {\mbox{tr}}

\def\sin {\mbox{sin}}
\def\PL #1 #2 #3 {Phys. Lett. {\bf#1} (#2) #3}
\def\NP #1 #2 #3 {Nucl. Phys. {\bf#1} (#2) #3}
\def\NPP #1 #2 #3 {Nucl. Phys. {\bf B} (Proc. Suppl.) {\bf#1} (#2) #3}
\def\ZP #1 #2 #3 {Z.~Phys. {\bf#1} (#2) #3}
\def\PR #1 #2 #3 {Phys. Rev. {\bf#1} (#2) #3}
\def\PP #1 #2 #3 {Phys. Rep.{\bf#1} (#2) #3}
\def\PRL #1 #2 #3 {Phys. Rev. Lett. {\bf#1} (#2) #3}
\def\PTP #1 #2 #3 {Prog. Theor. Phys. {\bf#1} (#2) #3}
\def\MPL #1 #2 #3 {Mod. Phys. Lett. {\bf#1} (#2) #3}
\def\IJM #1 #2 #3 {Int. J.~Mod. Phys. {\bf#1} (#2) #3}
\title{A Conclusive Test of Abelian Dominance Hypothesis 
       for Topological Charge in the QCD Vacuum\thanks{Talk presented 
       by S.~Sasaki at LATTICE 98, Boulder.}}

\author{S. Sasaki\address{Yukawa Institute for Theoretical Physics
(YITP), Kyoto University, %\\
Kyoto 606-8502, Japan}%
\thanks{Research Fellow of the Japan Society for the Promotion of Science.}
and 
O. Miyamura\address{Department of Physics, Hiroshima University, %\\
Higashi-Hiroshima 739-0046, Japan}}       
\begin{document}

%%%%%%%%%%%%%%%%%%%% Abstract %%%%%%%%%%%%%%%%%%%%%%%%
\begin{abstract}
We study the topological feature in the QCD vacuum
based on the hypothesis of abelian dominance.
The topological charge $Q_{\rm SU(2)}$ can be explicitly
represented in terms of the monopole current
in the abelian dominated system.
To appreciate its justification, we directly
measure the corresponding topological charge $Q_{\rm Mono}$, 
which is reconstructed only
from the monopole current and the abelian
component of gauge fields, by using the
Monte Carlo simulation on SU(2) lattice.
We find that there exists a one-to-one
correspondence between $Q_{\rm SU(2)}$ and $Q_{\rm Mono}$
in the maximally abelian gauge.
Furthermore, $Q_{\rm Mono}$
is classified by approximately discrete values.
\end{abstract}

% typeset front matter (including abstract)
\maketitle

%\section{Introduction}
A stimulating idea of the abelian gauge fixing was proposed 
by 't Hooft \cite{tHooft} and also independently by Ezawa-Iwazaki \cite{Ezawa}. 
After performing the partial gauge 
fixing to remain the abelian gauge degrees of freedom, one knows that 
the non-abelian gauge 
theory is reduced to the abelian gauge theory with magnetic monopoles. 
Such topological excitations would play an essential role on color
confinement \cite{Polikarpov}.
Furthermore, Ezawa and Iwazaki stressed the hypothesis of abelian 
dominance \cite{Ezawa}:
\begin{itemize}
   \item Only the abelian component 
         of gauge fields is relevant at a long-distance scale. 

   \item The non-abelian effects are mostly inherited by magnetic monopoles. 
\end{itemize}
They then demonstrated monopole condensation in a long-distance 
scale based on an argument about the ``energy-entropy balance'' on the 
monopole current \cite{Ezawa}.

The recent lattice QCD simulations show several evidences of monopole 
condensation \cite{Polikarpov}. This means that confinement could be regarded 
as the dual version of the superconductivity. Second important result from the 
lattice QCD simulation indicates that abelian dominance for 
some physical quantities, \eg the string tension \cite{Polikarpov} 
and the chiral condensate \cite{Miyamura}, 
is actually realized in the maximally abelian (MA) gauge.
In this gauge, at least, the abelian component could be 
the important dynamical degrees of freedom at a long-distance 
scale.
Here, an unavoidable question arises. 
{\it In the abelian dominated system, is it possible for the non-abelian
topological nature to survive?}
For such a question, Ezawa and Iwazaki have proposed a remarkable 
conjecture \cite{Ezawa}: once abelian dominance is postulated, 
the topological feature is preserved by the presence of monopoles.

Following the previous study of the fermionic zero-mode \cite{Sasaki}, 
we will show that the topological charge is explicitly represented 
in terms of monopoles in the lattice formalism 
by using the hypothesis of abelian dominance.
Finally, we intend to numerically confirm its justification by means 
of the Monte Carlo simulation in the MA gauge.
For simplicity, we restrict the discussion to the case of the SU(2) 
gauge group hereafter.

We first address the definition of the topological charge in the lattice
formalism. 
The naive and field-theoretical definition of the topological 
density is given by 
%
% eq. 1
%
\beqn
q(s) \equiv \frac{1}{2} \epsilon_{\mu \nu \rho \sigma} \tr
\left \{ P_{\mu \nu}(s) P_{\rho \sigma}(s) \right \}\;,
\label{Eq:su2top}
\eeqn
where the clover averaged SU(2) plaquette is defined as
$P_{\mu \nu}(s)\equiv\frac{1}{4}\left(\right.
U_{\mu \nu}(s) + U^{\dag}_{-\mu \nu}(s) 
+ U^{\dag}_{\mu -\nu}(s) + U_{-\mu -\nu}(s)
\left.\right)$ with the convenient notation for the SU(2) link variable as
$U_{-\mu}(s) = U^{\dag}_{\mu}(s-{\hat \mu})$.
The topological charge $Q_{\rm cont}$ 
is naively extracted from the summation of the previous defined topological 
density over all site as leading order in powers of the lattice spacing 
$a$:
%
% eq. 2
%
\beqn
Q_{L}=-\frac{1}{16\pi^2}\sum_{s}q(s) \simeq Q_{\rm cont} + {\cal 
O}(a^6)\;,
\eeqn
where $Q_{\rm cont}
=\frac{1}{16\pi^2}\sum_{s}\tr\left\{ a^{4} g^{2} G_{\mu \nu}(s) 
{}^{*}G_{\mu \nu}(s)\right\}$.

Here, the SU(2) link variable is expected to be U(1)-like as 
$U_{\mu}(s) \simeq u_{\mu}(s) \equiv \exp \left\{
i\sigma_{3}\theta_{\mu}(s) \right\}$, 
provided that the QCD vacuum is described as the 
abelian dominated system in a suitable abelian gauge.
The angular variable; 
$\theta_{\mu}(s) \equiv {\rm 
arctan}[{U^{3}_{\mu}(s)}/{U^{0}_{\mu}(s)}]$ is 
defined in the compact domain $[-\pi, \pi)$. 
In the abelian dominated system, we might consider 
the abelian analog of the topological density \cite{Bornyakov}
in stead of eq.(\ref{Eq:su2top}) as
%
% eq. 3
%
\barr
q_{_{\rm Abel}}(s) &=& \frac{1}{2} \varepsilon_{\mu 
\nu \rho \sigma} {\rm tr} \left\{ p_{\mu \nu}(s) p_{\rho 
\sigma}(s) \right\} \nonumber \\
&=& - \varepsilon_{\mu \nu \rho \sigma} {\cal S}_{\mu \nu}(s) 
{\cal S}_{\rho \sigma}(s) \;,
\earr
where $p_{\mu \nu}(s) \equiv \frac{1}{4}\sum_{i,j=0}^{1}u_{\mu \nu}
(s-i{\hat \mu}-j{\hat \nu})$ and ${\cal S}_{\mu 
\nu}(s)=\frac{1}{4}\sum_{i,j=0}^{1}\sin[\theta_{\mu \nu}
(s-i{\hat \mu}-j{\hat \nu})]$.

Our next aim is to discuss the expression of the abelian analog of the
topological density 
in the naive continuum limit $a \rightarrow 0$.  
This is because we need only the leading order term in powers 
of the lattice spacing to determine the corresponding topological charge.
Here, one may notice that the U(1) elementary plaquette $u_{\mu \nu}$ 
is a multiple valued function as the U(1) plaquette angle; $\theta_{\mu \nu}(s)\equiv 
\del_{\mu}\theta_{\nu}(s)-\del_{\nu}\theta_{\mu}(s)$. Hence, we 
divide into $\theta_{\mu \nu}$ into two parts as
%
% eq. 4
%
\beqn
\theta_{\mu \nu}(s) = {\bar \theta}_{\mu \nu}(s) + 2\pi n_{\mu 
\nu}(s)\;,
\eeqn
where ${\bar \theta}_{\mu \nu}$ is defined in the principal domain 
$[-\pi, \pi)$, which corresponds to the U(1) field strength. 
$n_{\mu \nu}$ can take the restricted integer; $0,\pm 1,\pm 2$. 
Taking the limit $a \rightarrow 0$, \ie ${\bar \theta}_{\mu \nu} 
\rightarrow 0$, we thus arrive at the following expression:
%
% eq. 5
%
\beqn
q_{_{\rm Abel}}(s) \approx - \varepsilon_{\mu \nu \rho \sigma}
{\bar \Theta}_{\mu \nu}(s){\bar \Theta}_{\rho \sigma}(s)\;,
\label{Eq:abeltop}
\eeqn
which is rewritten in terms of the U(1) field strength;
${\bar \Theta}_{\mu \nu} \equiv \frac{1}{4}\sum_{i,j=0}^{1}{\bar 
\theta}_{\mu \nu}(s-i{\hat \mu}-j{\hat \nu})$.

\noindent 

\begin{figure*}[hbt]
\hspace{0.0cm}
\begin{minipage}[hbt]{5cm}
\epsfysize = 4.8cm
\epsfbox{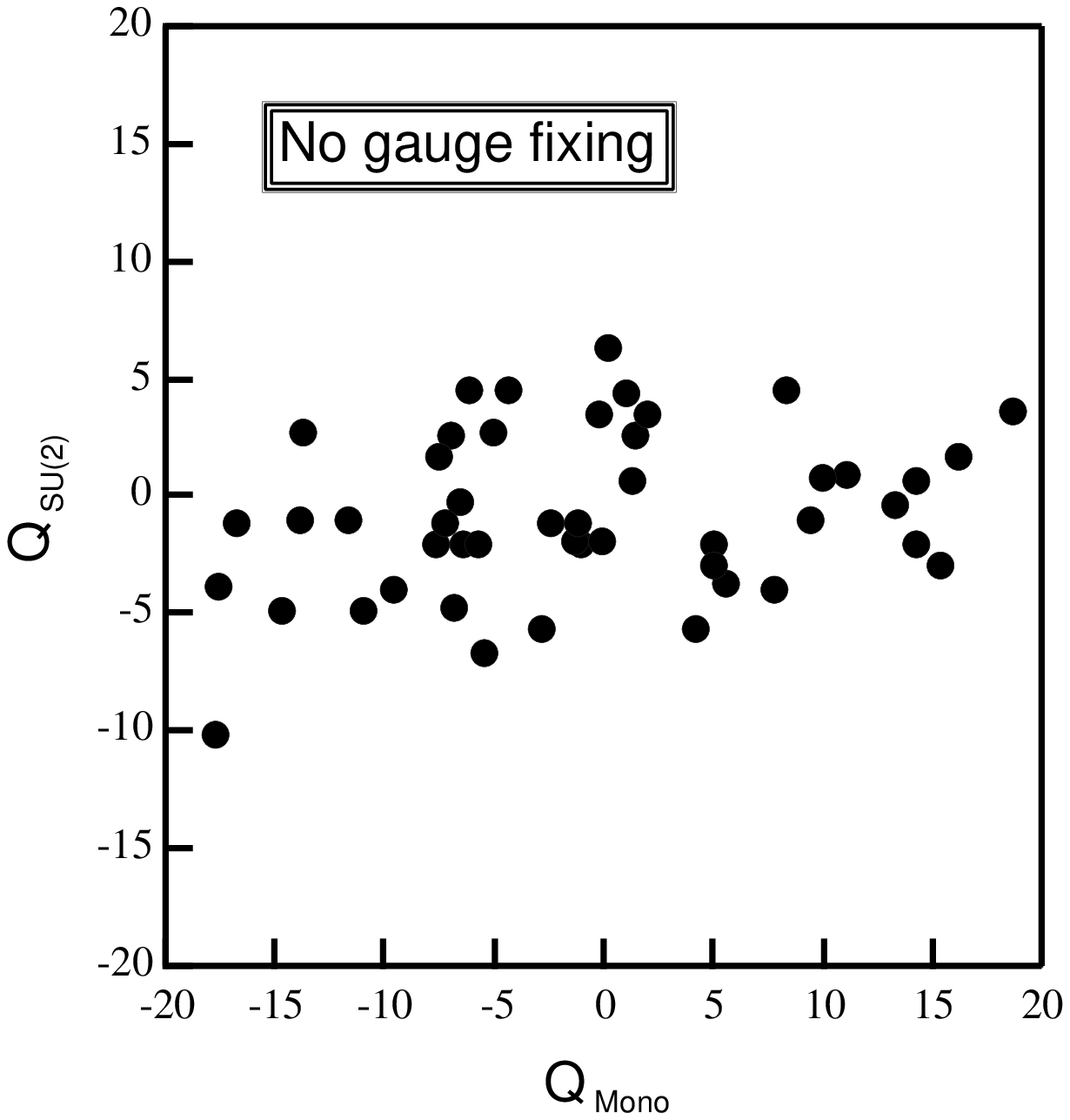}
%\framebox[50mm]{\rule[-21mm]{0mm}{43mm}}
\end{minipage}
\hspace{0.2cm}
\begin{minipage}[hbt]{5cm}
\epsfysize = 4.8cm
\epsfbox{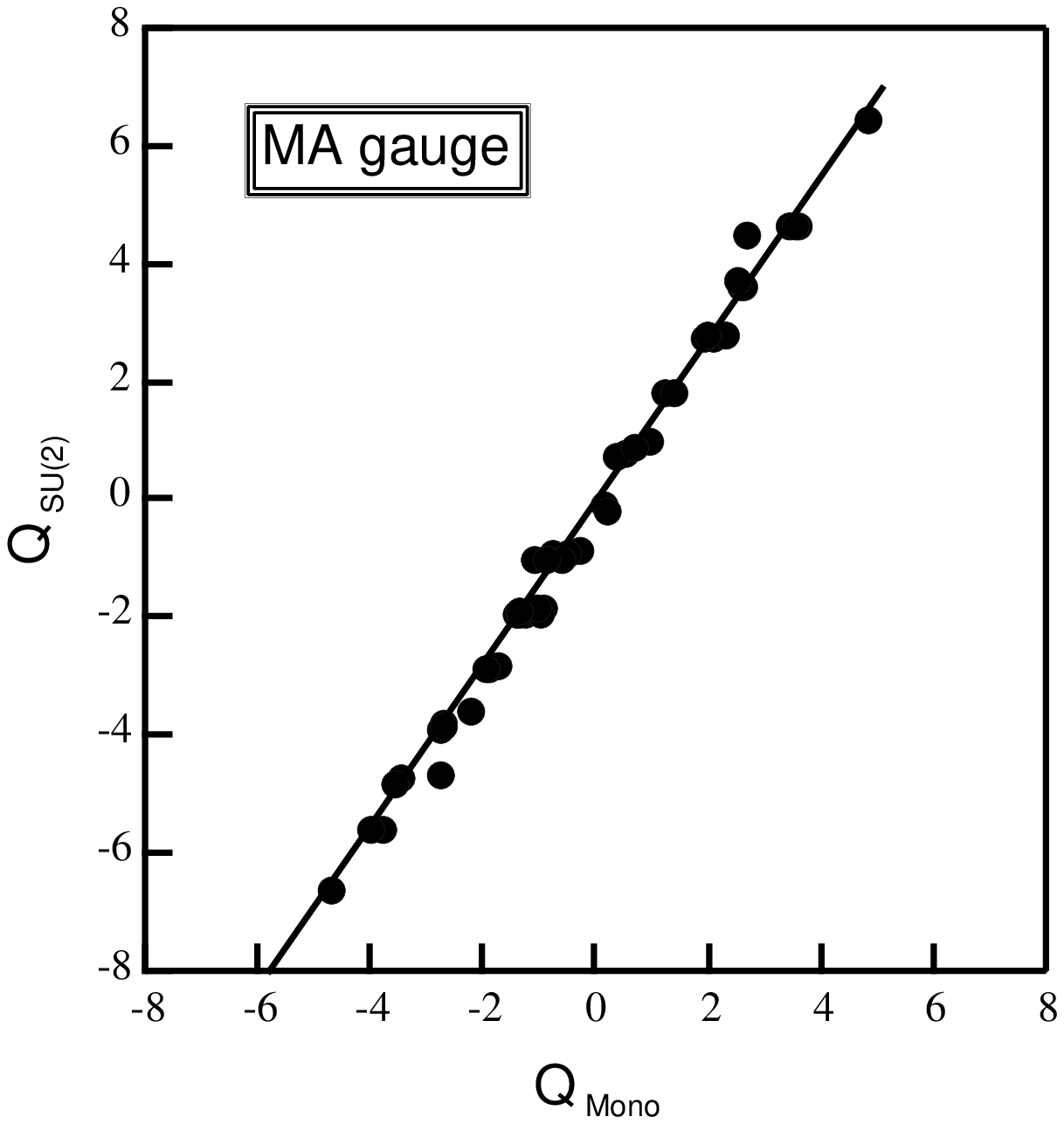}
%\framebox[50mm]{\rule[-21mm]{0mm}{43mm}}
\end{minipage}
\hspace{0.2cm}
\begin{minipage}[hbt]{5cm}
\epsfysize = 4.8cm
\epsfbox{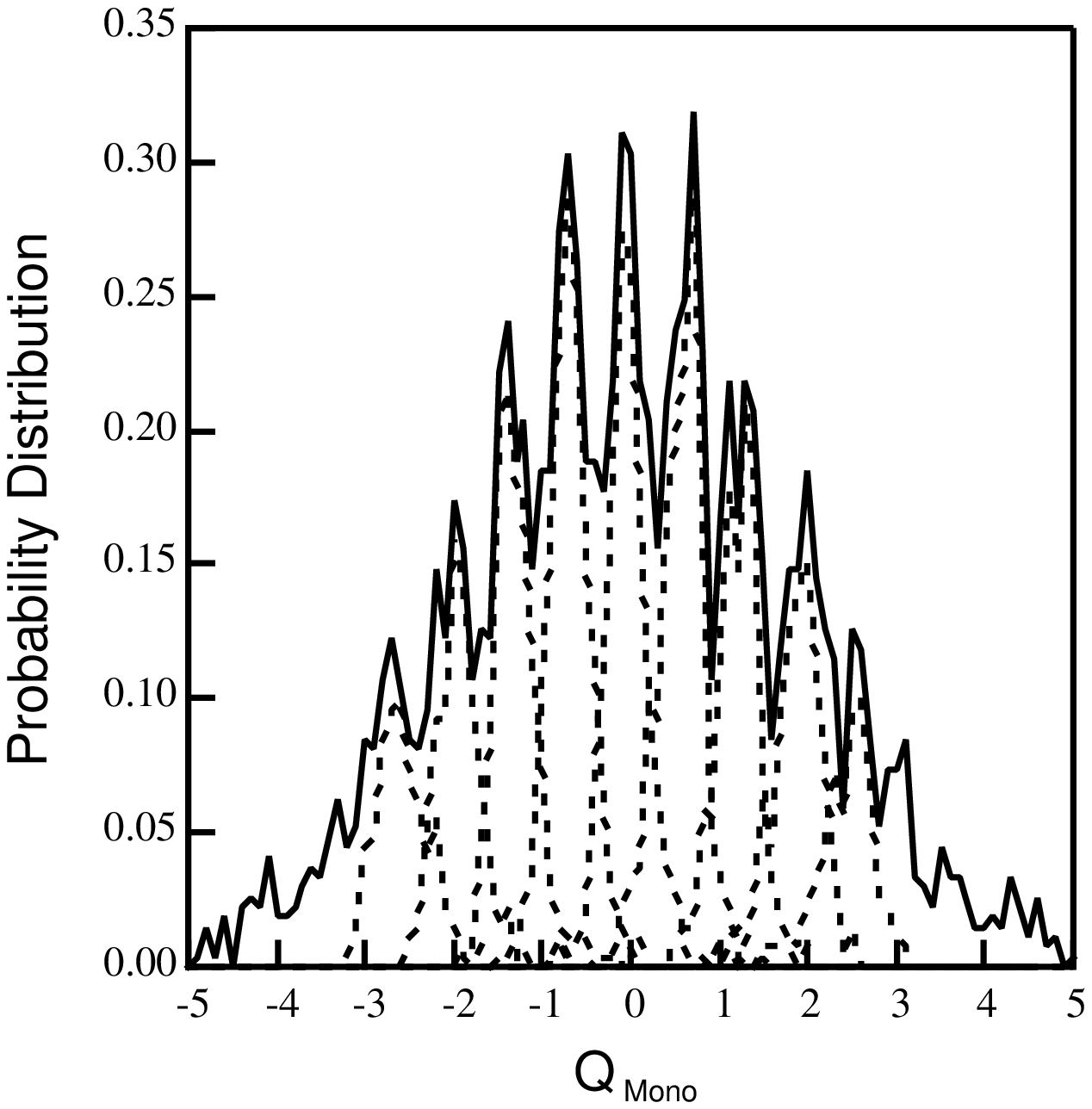}
%\framebox[50mm]{\rule[-21mm]{0mm}{43mm}}
\end{minipage}
\noindent
\vspace{0.0cm}\\

\hspace{0.3cm}
\begin{minipage}[hbt]{4.5cm}
Figure 1(a): Scatter plot of $Q_{\rm SU(2)}$ vs. $Q_{\rm Mono}$
in the case of no gauge fixing.
\end{minipage}
\hspace{0.7cm}
\begin{minipage}[hbt]{4.5cm}
Figure 1(b): Scatter plot of $Q_{\rm SU(2)}$ vs. $Q_{\rm Mono}$
in the maximally abelian (MA) gauge.
\end{minipage}
\hspace{0.8cm}
\begin{minipage}[hbt]{4.5cm}
Figure 2: Probability distribution of $Q_{\rm Mono}$ 
in the maximally abelian (MA) gauge.
\end{minipage}
\vspace{0.0cm}
\end{figure*}

Next, we intend to explicitly represent the topological charge in terms of 
monopoles. For the identification of monopoles, we follow
DeGrand-Toussaint's definition in the compact U(1) gauge theory 
\cite{DeGrand}. The monopole current is given by
%
% eq. 6
%
\beqn
k_{\mu}(s)
\equiv \frac{1}{4\pi}\epsilon_{\mu \nu \rho \sigma}\del_{\nu}
{\bar \theta}_{\rho \sigma}(s+{\hat \mu}) \;,
\eeqn
which denotes the integer-valued magnetic current, 
because of the Bianchi identity on the U(1) plaquette angle; 
$\epsilon_{\mu \nu \rho \sigma}\del_{\nu}\theta_{\rho \sigma}=0$
\cite{DeGrand}.

To show the explicit contribution of monopoles to the topological
charge, we introduce the dual potential 
${\cal B}_{\mu}$ satisfying the following equation \cite{Smit}:
%
% eq. 7
%
\beqn
\left(\Delta^2\delta_{\mu \nu}
-\Delta_{\mu}\Delta_{\nu} \right){\cal B}_{\nu}(s)
=-2\pi{\cal K}_{\mu}(s)\;,
\eeqn
where ${\cal K}_{\mu}(s)
\equiv \frac{1}{8} \sum_{i,j,k=0}^{1}k_{\mu}(s-i{\hat \nu}
-j{\hat \rho}-k{\hat \sigma})$ and 
$\Delta_{\mu}$ denotes the nearest-neighbor central difference 
operator.
It is worth mentioning that ${\cal K}_{\mu}$ satisfies the 
conservation law; $\Delta_{\mu}{\cal K}_{\mu}(s)=0$.
We can perform the Hodge decomposition on ${\bar \Theta}_{\mu \nu}$
with the dual potential ${\cal B}_{\mu}$ as
%
% eq. 8
%
\beqn
{\bar \Theta}_{\mu \nu}(s)=\Delta_{\mu}{\cal A}^{\prime}_{\nu}(s)
-\Delta_{\nu}{\cal A}^{\prime}_{\mu}(s)
+\varepsilon_{\mu \nu \rho 
\sigma}\Delta_{\rho}{\cal B}_{\sigma}(s),
\eeqn
where ${\cal A}^{\prime}_{\mu}$ is the Gaussian fluctuation 
\cite{Smit}.
After a little algebra,
we find the explicit contribution of monopoles to the
r.h.s of eq.\ref{Eq:abeltop} as
%
% eq. 9
%
\beqn
\varepsilon_{\mu \nu \rho \sigma}
{\bar \Theta}_{\mu \nu}(s){\bar \Theta}_{\rho \sigma}(s)
=16\pi {\cal A}^{\prime}_{\mu}(s) {\cal K}_{\mu}(s) + 
\cdot\cdot\cdot\;,
\eeqn
where the ellipsis stands for the total divergence, which will drop 
in the summation over all site.
Consequently, we arrive at the conjecture that 
{\it the topological feature is preserved by 
the presence of monopoles in the abelian dominated system}:
%
% eq. 10
%
\beqn
Q_{\rm cont}\simeq - \frac{1}{16\pi^2}\sum_{s}q_{_{\rm Mono}}(s)\;,
\eeqn
where $q_{_{\rm Mono}}(s)\equiv -16\pi {\cal A}^{\prime}_{\mu}(s)
{\cal K}_{\mu}(s)$.

%\section{Numerical Results}

Finally, we investigate above conjecture to justify 
by means of the Monte Carlo simulation.
We generate the gauge configurations by using the standard SU(2) Wilson 
action on a $16^4$ lattice at $\beta=2.4$.
First, we get the smoothed gauge 
configurations, which are eliminated undesirable fluctuations 
from the given Monte Carlo configurations through the naive 
cooling procedure. The realization of abelian dominance is established in
the MA gauge \cite{Polikarpov}. We then apply the smoothed gauge 
configurations after several cooling sweeps to the gauge transformation 
in order to impose the MA gauge condition.
We perform the Cartan decomposition on the gauge fixed 
SU(2) link variable and then obtain the U(1) field strength and the monopole 
current. Finally, we measure two types of the 
corresponding topological charge:
\begin{itemize}
   \item $Q_{\rm SU(2)}\equiv - \frac{1}{16 \pi^2} \sum_{s} q(s)$
   
   \item $Q_{\rm Mono}\equiv - \frac{1}{16 \pi^2} \sum_{s} q_{_{\rm 
   Mono}}(s)$
\end{itemize}

We show the scatter plots of $Q_{\rm SU(2)}$ vs. $Q_{\rm Mono}$
after 100 cooling sweeps in Fig.1; (a) no gauge fixing and (b) the MA 
gauge fixing.
Obviously, there 
are no any correlation between two topological charges in the case of 
no gauge fixing, where abelian dominance is not realized.
After the MA gauge fixing, a one-to-one correspondence between $Q_{\rm SU(2)}$ 
and $Q_{\rm Mono}$ reveals in the scatter plot.
In further detail, there is a small variance between $Q_{\rm SU(2)}$ and 
$Q_{\rm Mono}$. The slope of the scatter plot is not unity, but 
1.34 in Fig.1 (b).
In other words, it seems that the relation;
$Q_{\rm Mono}\approx 0.75 Q_{\rm SU(2)}$ is almost satisfied in the 
MA gauge on this data.
To discuss the topological feature on $Q_{\rm Mono}$, we show the 
probability distribution of $Q_{\rm Mono}$ in Fig.2 by using 3000 
independent configurations.
Several dotted lines correspond to partial contributions to the whole 
distribution, which are assigned to some integer value of $Q_{\rm SU(2)}$.
Several discrete peeks in Fig.2 tell us that $Q_{\rm Mono}$ is classified by 
approximately discrete values. Namely, it seems that monopoles almost inherit 
the topological nature.

In conclusion, we have studied the topological aspects of the QCD 
vacuum based on the hypothesis of abelian dominance.
We have found that the topological charge 
could be reconstructed from the monopole current 
and the abelian component of gauge fields if abelian dominance 
is realized. This indicates that the presence of monopoles preserves 
the non-abelian topological feature in the abelian dominated system.

%%%%%%%%%%%%%%%%%%%%%%%%%%%%

%%%%%%%%%%%%%%%%%%%%%%%%%%%%
\end{document}